\documentclass[12pt,draftclsnofoot,journal,onecolumn]{IEEEtran}
\usepackage{amsmath}
\usepackage{amssymb}

\usepackage{graphicx}
\usepackage{epstopdf}
\usepackage{amsmath}
\usepackage{array}
\newcommand{\PreserveBackslash}[1]{\let\temp=\\#1\let\\=\temp}
\newcolumntype{C}[1]{>{\PreserveBackslash\centering}p{#1}}
\newcolumntype{R}[1]{>{\PreserveBackslash\raggedleft}p{#1}}
\newcolumntype{L}[1]{>{\PreserveBackslash\raggedright}p{#1}}
\usepackage[usenames]{color}
\usepackage{colortbl,booktabs}
\usepackage{tabularx}
\usepackage{multicol}
\usepackage{booktabs}
\usepackage[Symbol]{upgreek}
\usepackage{bm}
\usepackage{cite}
\usepackage{balance}
\usepackage[boxed]{algorithm2e}
\usepackage{mathrsfs}
\usepackage{url}
\usepackage{subfigure}
\usepackage{threeparttable}
\usepackage{amsmath}
\usepackage{dcolumn}
\usepackage{multirow}
\usepackage{booktabs}
\usepackage{amsfonts}
\usepackage{amsthm,amsmath,amssymb}
\usepackage{mathrsfs}
\begin{document}
	\title{Antenna Array Diagnosis for Millimeter-Wave MIMO Systems\vspace{0 mm}}
	%
	%\author{\IEEEauthorblockN{Wenqian Shen, Linglong Dai, and Zhaocheng Wang}\vspace{-0mm}}
	\author{Siqi Ma,~\IEEEmembership{}
		Wenqian Shen,~\IEEEmembership{}
        Jianping An,~\IEEEmembership{}
        Lajos Hanzo,~\IEEEmembership{Fellow, IEEE}
		\thanks{S. Ma, W. Shen, J. An are with the Department of Information and Electronics, Beijing Institute of Technology, Beijing 100081, China (E-mails: sqma@bit.edu.cn; wshen@bit.edu.cn; an@bit.edu.cn). L. Hanzo is with the Department of Electronics and Computer Science, University of Southampton, Southampton SO17 1BJ, UK (e-mail: lh@ecs.soton.ac.uk).}
		
		%\thanks{A. Sayeed is with the Department of Electrical and Computer Engineering, University of Wisconsin, Madison, WI 53706, USA (email: akbar@engr.wisc.edu).}
		\vspace{-8mm}
	}
	\maketitle
\begin{abstract}
The densely packed antennas of millimeter-Wave (mmWave) MIMO systems are often blocked by the rain, snow, dust and even by fingers, which will change the channel's characteristics and degrades the system's performance. In order to solve this problem, we propose a cross-entropy inspired antenna array diagnosis detection (CE-AAD) technique by exploiting the correlations of adjacent antennas, when blockages occur at the transmitter. Then, we extend the proposed CE-AAD algorithm to the case, where blockages occur at transmitter and receiver simultaneously. Our simulation results show that the proposed CE-AAD algorithm outperforms its traditional counterparts.
	\end{abstract}
	
	\begin{IEEEkeywords}
		Millimter-Wave MIMO, antenna array diagnosis, cross-entropy.
	\end{IEEEkeywords}
	\IEEEpeerreviewmaketitle
	
\section{Introduction}\label{S1}
\IEEEPARstart{}{}Millimeter-Wave (MmWave) multiple-input multiple-output (MIMO) techniques have a high promise for next-generation wireless communication systems due to the abundance of bandwidth\cite{1,2}. The short wave-lengths of mmWave frequencies allow a large number of antennas to be packed in a compact physical size for providing a high beamforming gain for compensating the heavy path loss of mmWave signals. Unfortunately, these densely packed antennas are subject to blockages by humans and dust, snow or water drops\cite{4,5,6}. These blockages will inevitably change the array's geometry, causing a change of the composite channels \cite{16}. Therefore, it is necessary to diagnose the blocked/faulty antennas.

Hence several antenna array diagnosis (AAD) techniques have been proposed in\cite{7,9,10,11}. Explicitly, Bucci et al. \cite{7} studied both partial blockage as well as complete blockage, and then proposed a modified genetic algorithm for calculating the characteristic parameters of blocked antennas. Bucci et al. \cite{9} also proposed a method of analyzing the near-field data to detect the location of blocked antennas. However, the AAD methods proposed in \cite{7} and \cite{9} require a large number of measurements, which will substantially increase the time required for concluding the diagnosis. To solve this problem, sophisticated AAD techniques based on compressive sensing (CS) were proposed in \cite{10,11}. Specifically, the AAD problem was formulated as a sparse signal recovery problem by Migliore \cite{10}, which can be solved by CS algorithms. By contrast, upon exploiting the correlations between blocked antennas, a group-blockage diagnosis technique was proposed by Eltayeb et al.\cite{11}. However, this method cannot be used, when both the receive and transmit antennas are blocked at the same time.\\
\hspace*{0.36cm}Against this background, we propose a cross-entropy-inspired AAD method (CE-AAD) for detecting the location of blocked antennas and for calculating the corresponding characteristic parameters of blocked antennas. Specifically, our proposed CE-AAD exploits the correlations among the blocked antennas, where the required number of measurements is significantly reduced. Then, we extend the proposed CE-AAD method to the scenario, where blockages occur simultaneously both at the transmitter and receiver. Finally, our simulations show that the proposed CE-AAD method outperforms its traditional counterparts. \\

%Our simulations verify the analytical results.~
\emph{Notation}: We use the following notation throughout the paper. We let a, $\mathbf{a}$, $\mathbf{A}$ represent the scalar, vector, and matrix respectively. $\rm vec\{\cdot \}$ denotes the vectorization of a matrix. $\rm ivec\{\cdot \}$ denotes the invectorization of a vector.
$(\cdot)^{\rm{T}}$,
$(\cdot)^{\text{H}}$, and $(\cdot)^{-1}$ denote the transpose,
conjugate transpose, and inverse of a matrix, respectively. The operator $ \circ $ and $\otimes$ represent the Hadamard-product and Kronecker product, respectively. $\mathbf {1}_{{N} \times {M}}$ denotes all $ {1}$ matrix of size $N \times M$.

\section{System Model}\label{S2}
In this section, we consider a uniform planar array (UPA) of antennas at the transmitter with $N_x$ equally spaced elements along the x-axis, and $N_y$ equally spaced elements along the y-axis. Thus the total number of antennas at the transmitter is $N_{\rm{T}}=N_x\times N_y$ . The receiver is equipped with a single antenna.  At the $k$-th measurement $(k =1,2 \cdots K)$, the received signal $g_k$ can be described as
\begin{align}\label{eq_1} % eq1
g_k=\mathbf{f}^{\rm{T}}_k\cdot {\rm{vec}}(\mathbf{H}){s} +  {{n_k}},
\end{align}
where $\mathbf{H}\in \mathbb{C}^{N_x\times N_y}$ is the channel matrix, $\mathbf{f}_k\in \mathbb{C}^{N_T\times 1}$ is the transmit precoding (TPC) vector at the $k$-th measurement, and $s$ is the transmitted symbol, while ${n_k} \sim {CN}(0,{\delta}^2)$ is the additive noise and ${\delta}^2$ is the noise power. Considering the classical multi-path channel model, $\mathbf{H}$ can be written as \cite{17}
\begin{align}\label{eq_1} % eq1
\mathbf{H} = \sum\limits_{{\ell} = 1}^L {{\beta _{\ell}}\cdot \mathbf{A}({\theta _{\ell}},{\varphi _{\ell}})},
\end{align}
where $L$ is the number of propagation paths and $\beta _{\ell}$ is the complex gain of the $\ell$-th path. Furthermore, $\mathbf{A}({\theta _{\ell}},{\varphi _{\ell}}) \in \mathbb{C}^{N_x \times N_y}$ is the antenna array response, where $\theta _{\ell}$ is the elevation angle-of-departure (AoD) and $\varphi _{\ell}$ is the azimuth AoD. The $(m,n)$-th element of $\mathbf{A}({\theta _{\ell}},{\varphi _{\ell}})$ is given by ${\frac{1}{{\sqrt {{N_x}{N_y}} }}e^{j(m-1)\frac{{2\pi d_x}}{\lambda }\sin \theta _{\ell} \cos \varphi _{\ell}}}{e^{j(n-1)\frac{{2\pi d_y}}{\lambda }\sin \theta _{\ell} \sin \varphi _{\ell}  }}$ \cite{17}, where ${d_x}$ and ${d_y}$ are the antenna spacing along the x-axis and y-axis, $\lambda$ is the wavelength.

When the antenna array is blocked, the received signal at the $k$-th measurement is given by
\begin{align}\label{eq_3} % eq1
r_k = \mathbf{f}^{\rm{T}}_k\cdot(\mathbf{b} \circ {\rm{vec}}(\mathbf{H}))s + {n_k} {,}
\end{align}
where $\mathbf{b}\in {\mathbb{C}^{N_{\rm{T}}{\rm{ \times }}1}}$ and the $n$-th element of vector ${\mathbf{b}}$ can be defined as
\begin{align}\label{eq_4} % eq1
{b_n} = \left\{ {\begin{array}{*{20}{c}}
{{a_n}}\\
1
\end{array}\begin{array}{*{20}{c}}
{{\rm{if}}{\mkern 1mu} {\kern 1pt} {\rm{the}}{\mkern 1mu} {\kern 2pt} n{{ {\tiny{-}} {\rm{th}}}}{\mkern 1mu} {\kern 2pt} {\rm{antenna}}{\mkern 2mu} {\kern 2pt} {\rm{is}}{\mkern 1mu} {\kern 2pt} {\rm{blocked}}}\\
{{\rm{otherwise}}}
\end{array}} \right. ,
\end{align}
where ${a_n} = {\tau _n}{e^{i{\Psi _n}}}$ is the characteristic parameter of the blocked antenna $n$. Here ${\tau _n}$ is the absorption coefficient following the uniform distribution between 0 and 1. If the $n$-th antenna is completely blocked, ${\tau _n}$ is equal to 0. Otherwise, ${\tau _n}$ is a positive value less than 1. The variable ${\Psi _n}$ denotes the blockage scattering coefficient following the uniform distribution between 0 and $2\pi$ \cite{20}.

 Without loss of generality, we let the training symbol $s$ be equal to 1. Using (1) and (3), we have
\begin{align}\label{eq_5} % eq1
{y_k} = {r_k} - {g_k} = {\mathbf{f}_k^{\rm{T}}}\underbrace {(\rm{vec}(\mathbf{H}) \circ \mathbf{b} - vec(\mathbf{H}))}_\mathbf{q} + {\widetilde n_k},
\end{align}
where $\mathbf{q}=[q_1,q_2,\cdots, q_{N_{\rm{T}}}]^{\rm{T}} \in \mathbb{C}^{{N_{\rm{T}}}\times1}$ and ${\widetilde n_k}\sim CN(0,{\delta}^2)$. After \emph{K} measurements, we obtain
%\begin{align}
%\underbrace {
%{\begin{bmatrix}
%\rm{y}_1\\
%\rm{y}_2\\
%\vdots\\
%\rm{y}_K\\
%\end{bmatrix}}}_\mathbf{Y} =
%\underbrace {{\begin{bmatrix}
%{{{\rm{f}}_{1,1}}}&{{{\rm{f}}_{1,2}}}& \cdots &{{{\rm{f}}_{1,{N_T}}}}\\
%{{{\rm{f}}_{2,1}}}&{{{\rm{f}}_{2,2}}}& \cdots &{{{\rm{f}}_{2,{N_T}}}}\\
% \vdots & \vdots & \vdots & \vdots \\
%{{{\rm{f}}_{K,1}}}&{{{\rm{f}}_{K,2}}}& \cdots &{{{\rm{f}}_{K,{N_T}}}}
%\end{bmatrix}}}_F
%\underbrace {{\begin{bmatrix}
%{{q_1}}\\
%{{q_2}}\\
% \vdots \\
%{{q_{{N_T}}}}
%\end{bmatrix}}}_\mathbf{q} +
%{\begin{bmatrix}
%{{{\widetilde e}_1}}\\
%{{{\widetilde e}_2}}\\
% \vdots \\
%{{{\widetilde e}_{{N_T}}}}
%\end{bmatrix}}.
%\end{align}
\begin{align}
\mathbf{y}=\mathbf{Fq}+\widetilde {\mathbf{n}},
\end{align}
where $\mathbf{y}=[y_1,y_2,\cdots y_K]^{\rm{T}} \in \mathbb{C}^{K \times1}$  and $\mathbf{F}=[\mathbf{f}_1,\mathbf{f}_2,\cdots \mathbf{f}_K]^{\rm{T}} \in \mathbb{C}^{K\times N_{\rm{T}}}$. The TPC matrix $\mathbf{F}$ can be designed as a random matrix, which can be realized by 2-bit phase shifters, whose elements are randomly selected from $\{1+i, 1-i, -1-i, -1+i\}$ \cite{11}. Since the number of blocked antennas is usually small, the vector $\mathbf{q}$ is a sparse vector. Thus, the AAD problem becomes a sparse signal recovery problem.
Through the estimation of $\mathbf{q}$, which is denoted by $\hat{\mathbf{q}}$, the locations of blocked antennas can be recognized by finding the positions of non-zero elements of $\widehat {\mathbf{q}}$. Then, the characteristic parameters of blocked antennas  ${\tau _n}$ and ${\Psi _n}$ can be readily obtained as
${\tau _n} = \left| {\frac{{{{\widehat q}_{n}}}}{{{{{{h}}}_{n}}}} + 1} \right|$ ,${\Psi _n} = \angle (\frac{{{{\widehat q}_{n}}}}{{{{ {{h}}}_{n}}}} + 1)$, where ${n \in \mathcal{S}_{q}}$, which is the index set of non-zero elements in the vector $\mathbf{q}$. Furthermore, $\mathbf{ {\mathbf{h}}}=\rm{vec}(\mathbf{H})$ and ${ {h}_n}$ is the $n$-th element of $\mathbf{ {\mathbf{h}}}$. While $\widehat q_n$ denotes the $n$-th element of $\widehat {\mathbf{q}}$.
%\begin{align}\label{eq_7} % eq1
%{\tau _n} = \left| {\frac{{{{\widehat q}_{n \in S}}}}{{{{\widetilde A}_{n \in S}}}} + 1} \right| ,
%\end{align}
%\begin{align}\label{eq_8} % eq1
%{\Psi _n} = \angle (\frac{{{{\widehat q}_{n \in S}}}}{{{{\widetilde A}_{n \in S}}}} + 1) ,
%\end{align}
%where $\mathbf{ \widetilde A}=\rm{vec}(\mathbf{A})$, $\widetilde A_n$ is $n$th element of $\mathbf{ \widetilde A}$. $\widehat q_n$ denotes $n$th element of $\widehat {\mathbf{q}}$.

\section{Proposed CE-AAD Algorithm}\label{S3}
Let us now outline the CE-AAD algorithm by exploiting the correlations between the adjacent blocked antennas to detect the location of blocked antennas and calculate the corresponding characteristic parameters.

Since numerous antennas are packed in a compact physical size in mmWave MIMO system, several neighboring antenna elements might be blocked. This results in $\mathbf{Q}$ = {\rm ivec}($\mathbf{q}$) $\in \mathbb{C}^{N_x\times N_y}$ of (6) becoming a block-sparse matrix\cite{11}. An example of the block-sparse matrix $\mathbf{Q}$ is shown in Fig. 1.

%\begin{figure}[t]
%\center{\includegraphics[width=0.45\textwidth]{test4.eps}}
%\caption{(a) is an example of matrix $\mathbf{As}$ of 100 antennas with 12 blocked elements. (b) is an example of sparse structure of matrix $\mathbf{As}$ with 6 antennas 2 blocked antennas at receiver and 8 antennas 2 blocked antennas at transmitter. Yellow-block is blocked element and white-block is free element.}
%\label{ula}
%\end{figure}
\begin{figure}[t]
\center{\includegraphics[height=1.7in,width=2.65in]{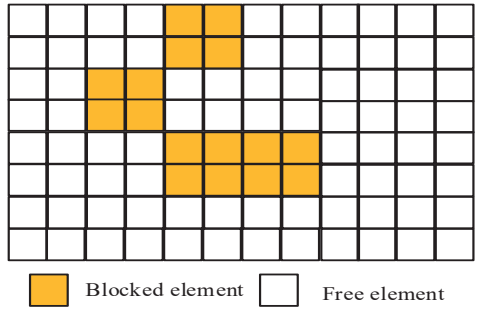}}
\caption{An example of the block-sparse matrix $\mathbf{Q}$. }
\label{ula}
\end{figure}
By exploiting the block-sparsity of matrix $\mathbf{Q}$, we propose a CE-AAD algorithm for recovering Q using a low number of measurements. Specifically, we use the vector $\mathbf{d}$ of size $N_{\rm{T}} \times 1$ to denote the positions of non-zero elements of vector $\mathbf{q}$, where the elements of $\mathbf{d}$ are selected from \{0,1\}. That is to say, ${q_n}=0$ means $d_n=0$ and ${q_n}\ne 0$ means ${d_n}=1$, where $d_n$ is the $n$-th element of $\mathbf{d}$. Thus, we can rewrite (6) as
%\begin{align}\label{eq_9} % eq1
%\underbrace {
%{\begin{bmatrix}
%\rm{y}_1\\
%\rm{y}_2\\
%\vdots\\
%\rm{y}_K\\
%\end{bmatrix}}}_Y\!=\! \underbrace { {\begin{bmatrix}
%{{{\rm{f}}_{1,1}}}&{{{\rm{f}}_{1,2}}}& \cdots &{{{\rm{f}}_{1,{N_T}}}}\\
%{{{\rm{f}}_{2,1}}}&{{{\rm{f}}_{2,2}}}& \cdots &{{{\rm{f}}_{2,{N_T}}}}\\
% \vdots & \vdots & \vdots & \vdots \\
%{{{\rm{f}}_{K,1}}}&{{{\rm{f}}_{K,2}}}& \cdots &{{{\rm{f}}_{K,{N_T}}}}
%\end{bmatrix}} }_F\underbrace {\ {\begin{bmatrix}
%{{d_1}\! \circ \!{q_1}}\\
%{{d_2} \!\circ \!{q_2}}\\
% \vdots \\
%{{d_{{N_T}}}\! \circ \!{q_{{N_T}}}}
%\end{bmatrix}} }_q\! +\!  {\begin{bmatrix}
%{{{\widetilde e}_1}}\\
%{{{\widetilde e}_2}}\\
% \vdots \\
%{{{\widetilde e}_{{N_T}}}}
%\end{bmatrix}} .
%\end{align}
\begin{align}
\mathbf{y}=\mathbf{F} ({\mathbf{q}} \circ {\mathbf d})+\widetilde {\mathbf{n}}.
\end{align}

Next, we propose the CE-AAD algorithm for estimating $\mathbf{d}$ and $\mathbf{q}$, as shown in Algorithm 1. Firstly, we initialize the algorithm by setting the probability matrix to ${\mathbf{P}}_{\rm B}^0 = {\frac{1}{2}} \times \mathbf{1}_{({N_x}/{N_{b_x}}) \times ({N_y}/{N_{b_y}})}$, where we assume that each block  is of size ${N_{b_x}} \times {N_{b_y}}$.
\begin{algorithm}
\caption{The proposed CE-AAD algorithm}
\KwIn{Precoding matrix $\mathbf{F}$; Measurement vector $\mathbf{y}$; Number of candidates $N_c$; Number of elites $N_e$; Number of iterations ${N_{iter}}$.}
$\mathbf{Initialization}$:
${\mathbf{P}}_{\rm B}^0= {1/2} \times {\mathbf{1}_{{({N_x}/{N_{b_x}})} \times ({N_y}/{N_{b_y}})}}$; $i=0$.

\For{$0\le i \le N_{iter}-1$}
{
 1. Randomly generate $N_c$ candidate vectors $\{ {\mathbf{d}_{n_c}}\} _{n_c = 1}^{N_c}$ according to ${\mathbf{P}_{\rm{B}}^{i}}$, where $\mathbf{d}_{n_c} \! \in\!  \{0, 1\}^{N_{\rm{T}} \times 1}$;

 2. ${\widehat {\mathbf{q}}^i}_{n_c} \!=\! {({\mathbf{F}_{n_c}^{\rm{H}}}\mathbf{F}_{n_c})^{ - 1}}{\mathbf{F}_{n_c}^{\rm{H}}}\mathbf{y}$;

 3. ${\zeta }^i_{n_c} = ||\mathbf{y} - \mathbf{F}_{n_c}{\widehat {\mathbf{q}}^i}_{n_c}|{|_2} + \epsilon ||{\mathbf{d}_{n_c}}|{|_0}$;

 4. Sort the objective function ${\zeta }^i_{n_c}$ in ascending order: $ {\zeta }^i_{d_{e,1}} \le {\zeta }^i_{d_{e,2}} \le\cdots \le{\zeta }^i_{d_{e,{N_{\rm{c}}}}}$;

 5. ${{\mathbf{P}_{\rm{B}}^{i + 1}}\! =\! \frac{1}{{N_e}}({\mathbf{C}_{\rm{B}}^{d_{e,1}}}\! + \!{\mathbf{C}_{\rm{B}}^{d_{e,2}}}\! + \! \cdots \!+ \!{\mathbf{C}_{\rm{B}}^{d_{e,N_e}}})}$;

 6. $i=i+1$;

}
\KwOut{Estimated value ${\widehat {\mathbf{q}}}$}
\end{algorithm}

\hspace*{-0.36cm}In Step 1, we divide the matrix $\mathbf{Q}$ into $({N_x}/{N_{b_x}}) \times ({N_y}/{N_{b_y}})$ blocks by exploiting the block-sparsity of $\mathbf{Q}$. We then define the probability matrix in the $i$-th iteration $\mathbf{P}_{\rm{B}}^i\in {\mathbb{C}^{N_x/{N_{b_x}} \times {N_y}/{N_{b_y}}}}$ , where the $(m,n)$-th element of $\mathbf{P}_{\rm{B}}^i$ denotes the probability that $\mathbf{C}_{\rm{B}}^{n_c}(m,n)=1$. Here, $\mathbf{C}_{\rm{B}}^{n_c}(m,n)$ is the $(m,n)$-th element of $\mathbf{C}_{\rm{B}}^{n_c} \in {\{0,1\}}^{({N_x}/{N_{b_x}})\times({N_y}/{N_{b_y}})}$, representing whether the $(m,n)$-th block of $\mathbf {Q}$ is a non-zero block. We generate $N_c$ candidates $\{ {\mathbf{C}_{\rm{B}}^{n_c}}\} _{n_c = 1}^{N_c }$ based on the probability matrix $\mathbf{P}_{\rm{B}}^i$. Then, we vectorize $\mathbf{C}_{n_c}$ as $\mathbf{d}_{n_c}= {\rm{vec}}(\mathbf{C}_{n_c}) \in \mathbb{R}^{N_{\rm{T}} \times 1}$, where $\mathbf{C}_{n_c}= \mathbf{C}_{\rm{B}}^{n_c} \otimes \mathbf {1}_{{N_{b_x}} \times {N_{b_y}}} \in \mathbb{R}^{N_x \times N_y} $. We also define the index set $\mathcal{D}_{n_c}$ for recording the indices of non-zero elements in vector $\mathbf{d}_{n_c}$.
In Step 2, we use the least squares (LS)\cite{21} algorithm to compute ${\widehat {\mathbf{q}}^i}_{n_c}\in {\mathbb{C}}^{N_{n_c}\times 1}$, which is the estimate of the non-zero elements of $\mathbf{q}$ associated with the $n_c$-th candidate ${\mathbf d}_{n_c}$ in the $i$-th iteration, where $N_{n_c}$ is the number of none-zero elements in candidate ${\mathbf{ d}_{n_c}}$. We then define a submatrix $\mathbf{F}_{n_c} \in {\mathbb{C}}^{K \times N_{n_c}}$, which is composed of the column vectors of matrix $\mathbf F$ with indices belonging to $\mathcal{D}_{n_c}$.
In Step 3, we define an objective function as  ${\zeta }^i_{n_c} = ||\mathbf{y} - \mathbf{F}_{n_c}{\widehat {\mathbf{q}}^i}_{n_c}|{|_2} + \epsilon ||{\mathbf{d}_{n_c}}|{|_0}$, where $||{\mathbf{d}_{n_c}}|{|_0}$ is the ``$\mathcal{L}0$" norm of $\mathbf{d}_{n_c}$, i.e. the number of none-zero elements in $\mathbf{d}_{n_c}$, while $\epsilon$ is a super-parameter \cite{18} \cite{19}, which is set to 0.6 for our simulations.
In Step 4, we sort ${\zeta }^i_{n_c}$ in an ascending order and retain the first $N_e$ elements. Then, we select the corresponding $N_e$ candidates $\mathbf{d}_{n_c}$  as elites and record their indices as $\{d_{e,1} ,d_{e,2}, \cdots ,d_{e,N_e}\}$.
In step 5, we update the probability matrix as ${{\mathbf{P}_{\rm{B}}^{i + 1}}\! =\! \frac{1}{{N_e}}({\mathbf{C}_{\rm{B}}^{d_{e,1}}}\! + \!{\mathbf{C}_{\rm{B}}^{d_{e,2}}}\! + \! \cdots \!+ \!{\mathbf{C}_{\rm{B}}^{d_{e,N_e}}})}$.
We then repeat this procedure until the maximum number of iterations $N_{iter}$ is reached.

The traditional cross-entropy (CE) algorithm \cite{14} ignores the block-sparsity of matrix $\mathbf{Q}$. In contrast to the CE algorithm, the proposed CE-AAD algorithm generates ${\mathbf{d}}_{n_c}$ by exploiting the block-sparsity of $\mathbf{Q}$ in Step 1. The corresponding candidate matrix $\mathbf{C}_{n_c}$ exhibits a block-structure. In this way, the CE-AAD algorithm achieves better performance of recovery and get faster convergence.
Furthermore, when complete blockages occur, we have $a_n=0$. Then, the proposed algorithm will be further simplified, since Step 2 and Step 3 can be combined. The objective function becomes ${\zeta }^i_{n_c} = ||\mathbf{y} - \mathbf{F}{\mathbf{d}_{n_c}}|{|_2} + \epsilon ||{\mathbf{d}_{n_c}}|{|_0}$. Hence it is not necessary to estimate the value of ${\widehat {\mathbf{q}}^i}_{n_c}$ by exploiting the LS algorithm. After ${N_{iter}}$ iterations, $\mathbf{d}_{d_{e,1}}$ is the estimated value of $\mathbf {q}$.
\section{Joint CE-AAD}\label{S4}
In the previous section, we proposed the CE-AAD for the case where blockages only occur at the transmitter. In this section, we extend the proposed CE-AAD algorithm to the case where blockages simultaneously occur both at the transmitter and receiver.

We consider the uniform linear arrays (ULAs) of antennas at both the transmitter and receiver with $N_{\rm t}$ transmit and $N_{\rm r}$ receive antennas\footnote{The proposed technique can be easily extended to the UPA scenario.}, respectively.  Similar to (5), at the $k$-th measurement, we have
\begin{figure}[t]
\center{\includegraphics[height=1.9in,width=2.7in]{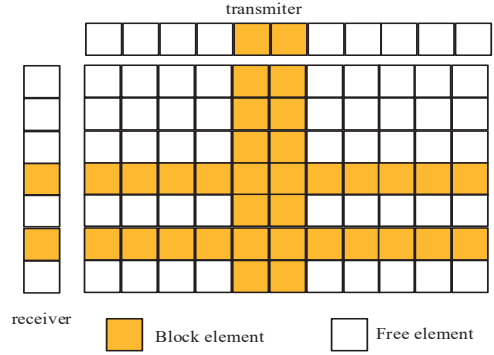}}
\caption{An example of sparse structure of matrix $\mathbf{Q}$.}
\label{ula}
\end{figure}

%\begin{align}\label{eq_13} % eq1
%\begin{array}{l}
%\begin{scriptsize}
%{\mathbf{w}_k}^{\rm{H}}\underbrace{\!(({\mathbf{b_r}}\! \circ \!{\mathbf{a_r}}(\theta ,\varphi )){{({\mathbf{b_T}}\! \circ \!{\mathbf{a_T}}(\theta ,\varphi ))}^{\rm{H}}}{ - }({\mathbf{a_r}}(\theta ,\varphi )){{({\mathbf{a_T}}\!(\theta ,\varphi ))}^{\rm{H}}})}_{{\mathbf{A_s}}}{{\mathbf{f}}_k} \!+\!\! \sqrt p {\widetilde e_k}.
%\end{scriptsize}
%\end{array}
%\end{align}	

\begin{align}\label{eq_13} % eq1
{y_k} = {r_k} - {g_k} = \mathbf{w}_k^{\rm{H}}\underbrace {(\mathbf{H} \circ \mathbf{B} - \mathbf{H})}_{\mathbf{Q}}{\mathbf{f}_k} + {\tilde n_k},
\end{align}	
where ${\mathbf{w}_k}\in {\mathbb{C}^{{N_{\rm r}} \times 1}}$ denotes the combining vector at the receiver and ${\mathbf{f}_k}\in {\mathbb{C}^{{N_{\rm t}} \times 1}}$ represents the TPC vector at the transmitter. Furthermore, $\mathbf{B}=\mathbf{b}_{\rm r}\mathbf{b}_{\rm t}^{\rm{T}}$, where $\mathbf{b}_{\rm r} \in {\mathbb{C}^{{N_{\rm r}} \times 1}} $ and $\mathbf{b}_{\rm t} \in {\mathbb{C}^{{N_{\rm t}} \times 1}}$ denote the blockage coefficients at the receiver and transmitter, which have similar definitions as $\mathbf{b}$ in (3), while $\mathbf{H} \in {\mathbb{C}^{N_{\rm r} \times N_{\rm t}}}$ is the channel matrix of
\begin{align}\label{eq_13} % eq1
\mathbf{H} = \sum\limits_{\ell = 1}^L {{\beta _\ell}{\mathbf{a}_{\rm r}}({\theta _\ell^r})} {\mathbf{a}_{\rm t}^{\rm H}}({\theta _\ell^t}),
\end{align}	
where $L$ is the number of propagation paths and $\beta _l$ is the complex gain of the $\ell$-th path. Furthermore, $\mathbf{a}_{\rm r}({\theta _\ell^r}) \in {\mathbb C}^{N_{\rm r} \times 1}$ and $\mathbf{a}_{\rm t}({\theta _\ell^t})\in {\mathbb C}^{N_{\rm t} \times 1}$ denote the antenna array responses at the receiver and transmitter, respectively. The steering vector $\mathbf{a}_{\rm x}({\theta _\ell^{\rm x}})$ can be expressed as $[1,{e^{ - j2\pi \frac{d}{\lambda }\sin (\theta _\ell^{\rm x} )}},\cdots,$ ${e^{ - j2\pi \frac{{({N_{\rm x}} - 1)d}}{\lambda }\sin (\theta _\ell^{\rm x})}}]^{\rm T}$, where ${\rm x} \in \{\rm t,r\}$ and $d$ is the antenna spacing. Vectorizing the matrix $\mathbf{Q}  \in {\mathbb{C}^{N_{\rm r} \times N_{\rm t}}}$, (8) can be rewritten as
\begin{align}\label{eq_14} % eq1
{y_k} = \underbrace {({{\mathbf{f}}}_k^{\rm{T}} \otimes {\mathbf{w}}_k^{\rm{H}})}_{{\mathbf{u}}_k}\underbrace {\rm{vec}({\mathbf{Q}})}_\mathbf{q} + {\widetilde n_k},
\end{align}
where the matrix $\mathbf{Q}$ has a sparse structure, as shown in Fig. 2. Therefore, after $K$ measurements, we have
\begin{align}
\mathbf{y}=\mathbf{Uq}+{\widetilde {\mathbf n}},
\end{align}	
where $\mathbf y = [y_1,y_2,\cdots, y_K]^{\rm T} \in \mathbb C^{K \times 1}$ and $\mathbf U=[{{\mathbf u}_1^{\rm T},{\mathbf u}_2^{\rm T},\cdots,{\mathbf u}_K^{\rm T}}]^{\rm T} \in \mathbb C^{K \times N_rN_t}$. The joint antenna array diagnosis becomes a sparse signal recovery problem.
\begin{figure*}[t]\centering\subfigure[] {\includegraphics[height=2.1in,width=2.25in,angle=0]{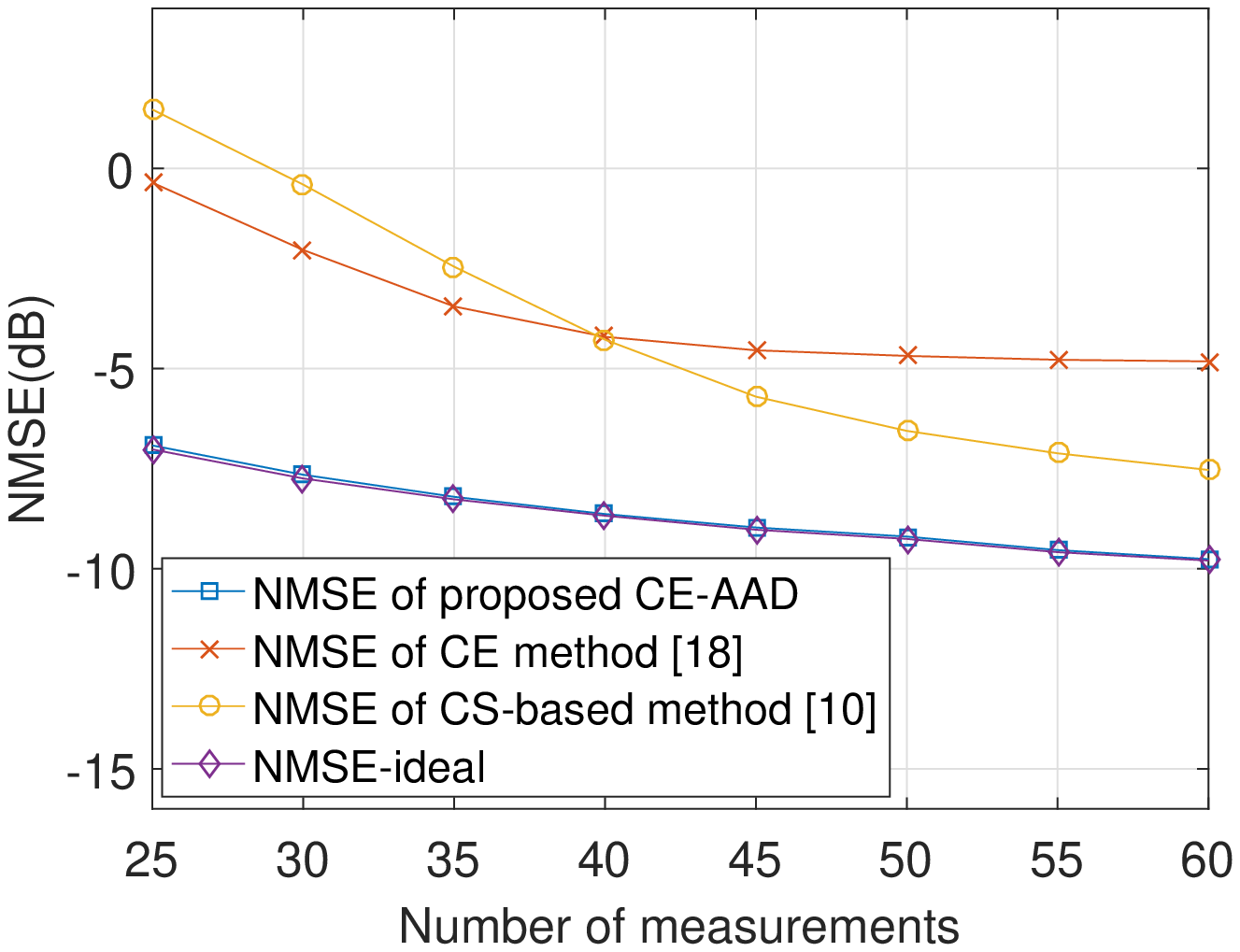}}\subfigure[] {\includegraphics[height=2.1in,width=2.25in,angle=0]{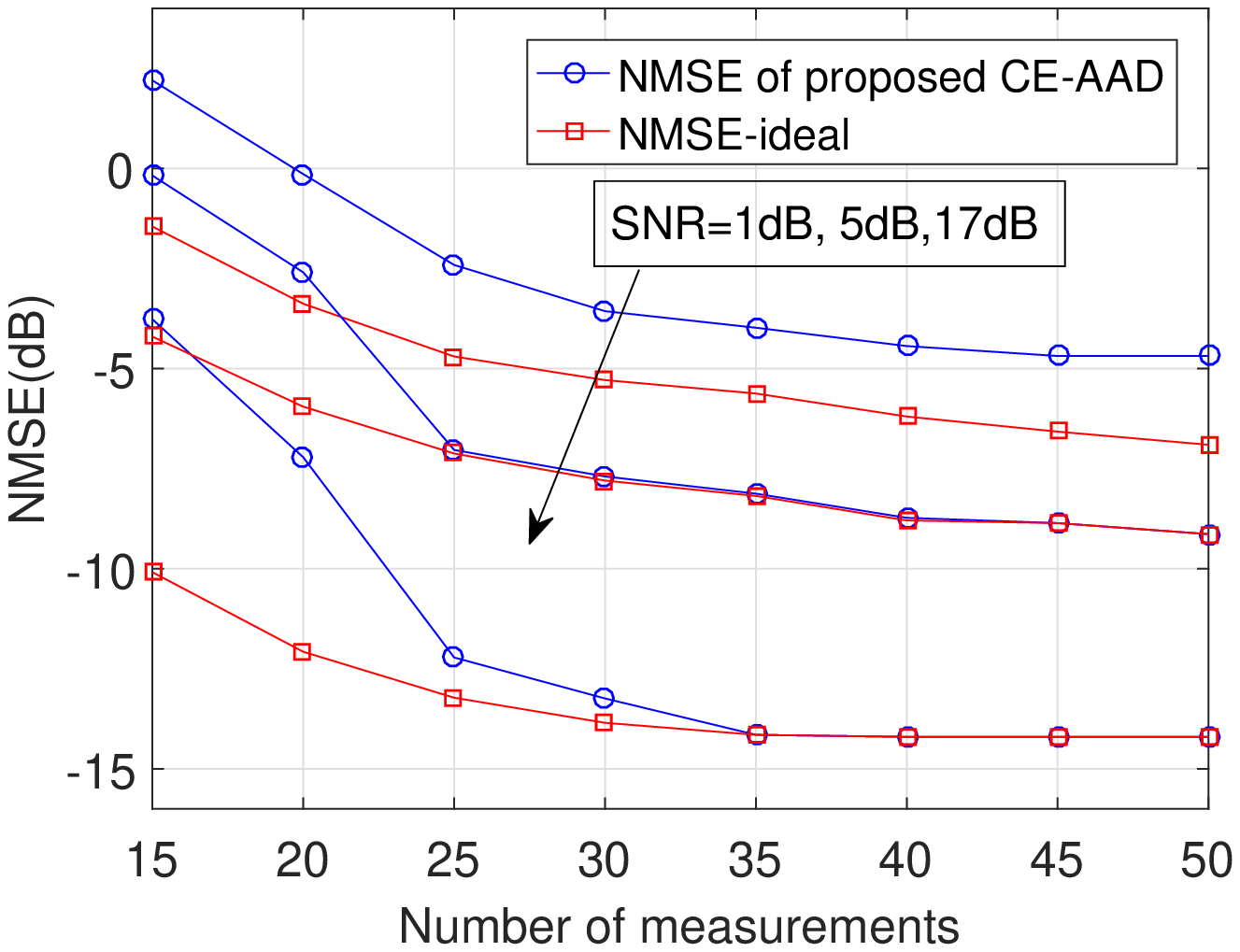}}\subfigure[] {\includegraphics[height=2.1in,width=2.25in,angle=0]{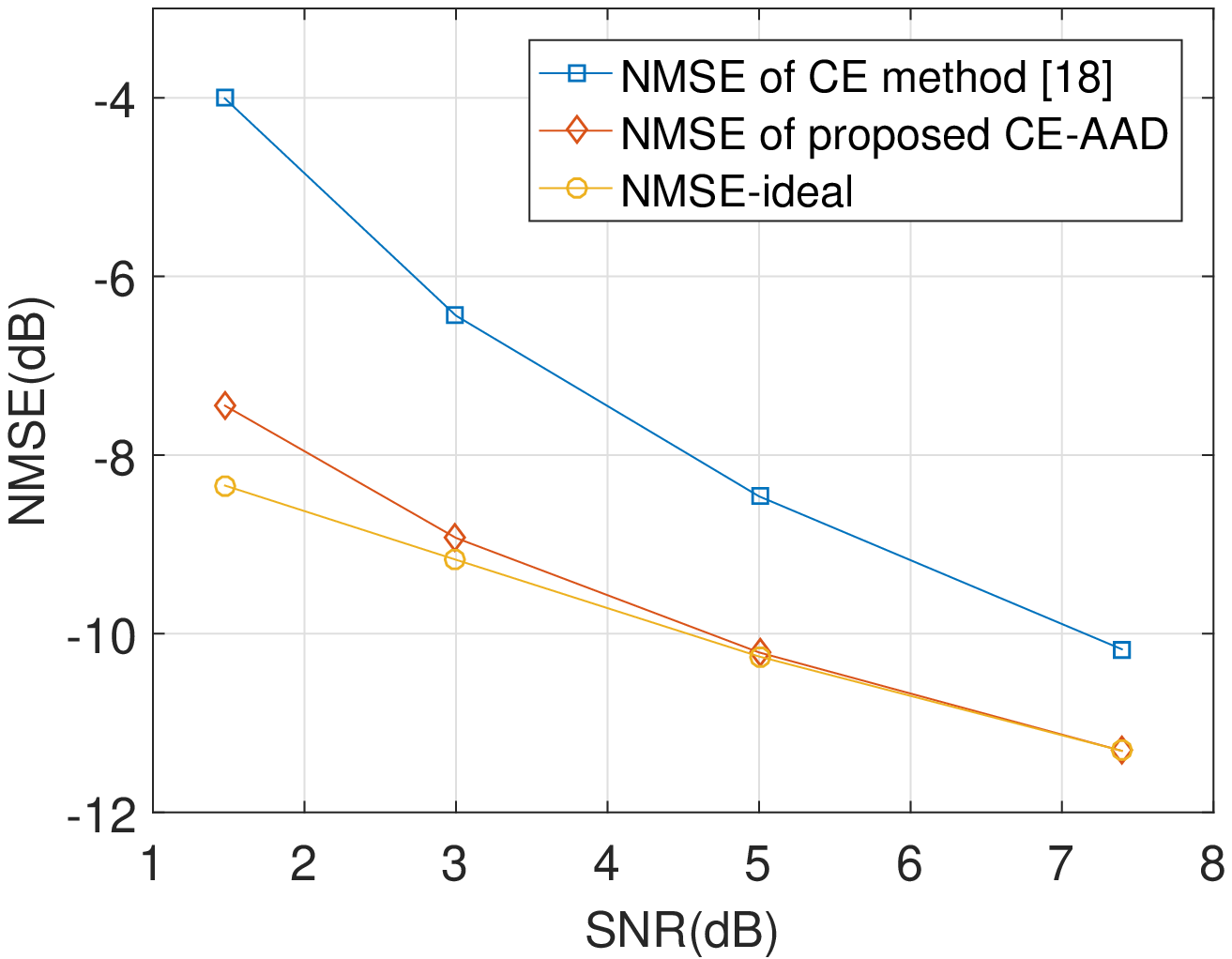}}\caption{NMSE performance comparison: (a) and (b) show the NMSE comparison when blockages occur at the transmitter. (c) shows the NMSE comparison when blockages occur at both the transmitter and receiver.}\label{fig5}\end{figure*}
Since $\mathbf Q$ has different sparse structures, the CE-AAD algorithm of this section is different from that in Section III. The main difference is reflected in Step 1 and Step 5.
In Step 1, we randomly generate $N_c$ candidate vectors $\mathbf{d}^{n_c}_{\rm t} \in {\{0,1\}^{{N_{\rm{t}}} \times 1}}$ according to the probability vector $\mathbf{p}_{\rm t}^i \in {\mathbb{R}}^{N_t \times 1}$ as well as $N_c$ candidate vectors ${\mathbf{d}}^{n_c}_{\rm r} \in {\{0,1\}^{{N_{\rm{r}}} \times 1}}$ according to the probability vector $\mathbf{p}_{\rm r}^i \in {\mathbb{R}}^{N_r \times 1}$. Then we have $N_c$ candidates $\mathbf{d}_{n_c} = \rm{vec}(\mathbf{C}_{n_c})$, where $\mathbf{C}_{n_c}(m,n) = \mathbf{d}^{n_c}_{\rm r}(m) \oplus \mathbf{d}^{n_c}_{\rm t}(n) $ denotes the $(m,n)$-th elements of $\mathbf{C}_{n_c}$. The operator $\oplus$ denotes the `or' operation. In Step 5, we use ${N_e}$ elites ${\mathbf{d}}^{d_{e,k}}_{\rm r}$ and ${\mathbf{d}}^{d_{e,k}}_{\rm t}$ $(k=1,2,\cdots N_e)$ to update the probability $\mathbf{p}^{i+1}_{\rm r}$ and $\mathbf{p}_{\rm t}^{i+1}$, where $d_{e,k}$ has the same definition as in Section III. After obtaining the estimation of $\mathbf{q}$, we can identify the locations of blocked antennas and calculate the corresponding characteristic parameters.
\section{Simulation Results}\label{S5}
In this section, we characterize the performance of the proposed CE-AAD algorithm. We adopt the ray-based mmWave channel model of \cite{15}. The system parameters used are as follows: $d=\frac{d_x}{\lambda}=\frac{d_y}{\lambda}=\frac{1}{2}$, $N_c=400$, $N_e=50$, $N_{iter}=20$, ${\epsilon = 0.6}$, ${\tau}_n \sim U(0,1)$, and ${{\Psi}_n \sim U(0,2\pi)}$. The number of dominant paths $L$ is 10, $\beta _\ell$ obeys the Gaussian distribution with zero mean and unit variance, while $\theta_\ell, \varphi_\ell, \theta_\ell^r, \theta_\ell^t$ are randomly chosen from $[-\pi/2,\pi/2]$. We assume that the blockage probability of antennas $p_b$ is 0.1. We simulate the normalized mean square error (NMSE) of $\mathbf b$ and $\mathbf{B}$ estimated both by the proposed CE-AAD method and by a pair of traditional AAD methods [10], \cite{13}.

In Fig. 3 (a) and (b), we consider a single free antenna at the receiver and a UPA having $10\times 10$ antennas at the transmitter. Fig. 3 (a) shows the NMSE comparison of different algorithms vs. the number of measurements when the signal-to-noise ratio (SNR) $\frac{1}{{\delta}^2}$ is 5dB. NMSE-ideal represents the performance upper-bound, which is achived when the exact locations of blocked antennas are known. We observe that the proposed algorithm outperforms other algorithms.In Fig. 3 (b), we compare the NMSE performance of the proposed algorithm and NMSE-ideal at different SNRs. We observe that the proposed algorithm achieves near-ideal performance using only a few measurements.

In Fig. 3 (c), we consider the case where blockages occur at both the transmitter and receiver, which rely on ULAs having 10 elements. Fig. 3 (c) shows our NMSE performance comparison vs. the SNRs, when the number of measurements is 50. We observe that the proposed joint CE-AAD algorithm has a good performance.

\section{Conclusions}\label{S5}
In this paper, we studied the AAD problem of mmWave MIMO systems. By exploiting the correlations between adjacent blocked antennas, we proposed the CE-AAD algorithm for identifying the locations and characteristic parameters of blocked antennas when blockages occurs at the transmitter. Then, we extended the proposed CE-AAD algorithm to the case, where blockages occur both at the transmitter and receiver simultaneously. Our simulation results verify that the proposed method achieves near-optimal performance, approaching that of the genie solution having explicit knowledge of the blocked antenna indices.
%	\newpage
	\bibliographystyle{IEEEtran}
	\bibliography{IEEEabrv,Refference}
	
\end{document}